\documentclass[prb]{revtex4}

\usepackage{graphicx}
\usepackage{dcolumn}
\usepackage{bm}

\begin{document}

\title{A proposal for a conditional definition of the Swap Gate}
\author{P. C. Garc\'{\i}a Quijas}
\affiliation{Universidad de Guanajuato\\
Loma del Bosque No. 113, Col. Lomas del Campestre,\\
C.P. 37150, Le\'{o}n, Gto. M\'{e}xico }
\author{L. M. Ar\'{e}valo Aguilar}
\email{larevalo@foton.cio.mx}
\homepage{http://www.cio.mx}
\affiliation{Centro de Investigaciones en \'{O}ptica, A. C.\\
Loma del Bosque No. 115, Col. Lomas del Campestre,\\
C.P. 37150, Le\'{o}n, Gto. M\'{e}xico }
\date{\today}
\textsc{Note}{:These authors contributed equally to this work}
\begin{abstract}
In order to realize quantum logical operations, Quantum Computation
(QC) requires that its basic tools and concepts obey the laws of
physics. One of the fundamental concepts in QC is the conditional
quantum dynamics \cite{barenco}, some times called
controlled-unitary operation \cite{kumar}, which is established by
the conditional "If-Then" sentence. The best know example is the
c-not gate, which operates on the
computational basis as follows: \textit{If the control qubit is set to $%
|1\rangle$, then apply the single qubit quantum NOT gate on the
target. Otherwise, if the control qubit is set to $|0\rangle$,
then the target qubit is unchanged}. This gate represents the
paradigm for the conditional quantum dynamics, where the flipping
of the target qubit is conditioned to the state of the control
qubit. Other gates have been defined in a similar way of
conditional evolution; for instance, the control phase gates
\cite{paulo1}. However, to the best of our knowledge, such
conditional quantum dynamics has not been yet used to define the
SWAP gate. Here we propose a possible conditional definition, in
the form of If-Then sentence, to construct a SWAP gate in the case
of two qubits. This definition suggests a classification in two
classes, which depends upon the number of qubits that have to
undergo a conditional quantum dynamics.
\end{abstract}

 \maketitle




\section{\label{sec:level1}Introduction}

The SWAP gate on two qubits is usually defined as
\cite{feynman,barenco,nielsen,deutsch}:
\begin{equation}
\hat{U}_{SWAP}\left\vert a\right\rangle _{1}\left\vert
b\right\rangle _{2}=\left\vert b\right\rangle _{1}\left\vert
a\right\rangle _{2},  \label{1}
\end{equation}%
where the principal task of the SWAP operation is to exchange the
state of the first qubit to second qubit, and vice versa. Then,
there has to be a physical process that makes the exchange of
states. This phenomenon is analogous to the classical example of a
perfect elastic collision between two particles of equal masses,
where exchange of momentum occurs. Therefore, the physical process
involved in the SWAP operation defined by equation (\ref{1}) is
the interchange of states of a composite system. It is worthwhile
mentioning that the SWAP operation is one of the most useful gate
in quantum computation. Its usefulness range from establishing the
universality of two qubit gates \cite{deutsch2}, programmable gate
arrays \cite{nielsen-a}, quantum teleportation
\cite{vaidman,vaidman2}, the basic implementation of other quantum
gates using SWAP operations \cite{loss,fan,brassard,eckert} (the
corresponding implementation have been realized with the
$\sqrt{SWAP}$) to the construction of optimal quantum circuits
\cite{vatan}.

On the other hand, as it is well known, three CNOT gates can
implement the SWAP gate
\cite{feynman,nielsen,barenco,abarenco,linden,fiorentino,vatan,fan,hardy}.
From this result there seem to be a different realization of the
SWAP operation, i. e. a realization in terms of a flipping
process. Therefore we can ask the following question: Is there a
quantum gate definition, which, by using both conditional
evolution and flipping process, produces the swapping of two
unknown qubits? The answer to this question is affirmative. To
show this case, let us proceed as follows.

First, let us define the SWAP gate, $\hat{U}_{SWAP}$ , as:
\begin{quote}
\textsf{\textsc{Definition 1}. If the states of a two parties
system are not equal, then apply the single qubit quantum NOT
gate, $\hat{U}_{NOT}$, on each subsystem. Otherwise, left them
unchanged.}\end{quote}
Where the states of a two parties system
are equal in the case $|0\rangle_1|0\rangle_2$ or
$|1\rangle_1|1\rangle_2$; that is, both in the ground state or
both in the excited state. The single qubit quantum NOT gate flips
the basic computational states of single qubits:
$\hat{U}_{NOT}|0\rangle=|1\rangle$ and
$\hat{U}_{NOT}|1\rangle=|0\rangle$ \cite{nielsen}.

\textsc{Definition 1} is a conditional sentence of the form
"if-then", which involves a crucial difference between both
definitions of the SWAP gate. In the case of definition given by
equation (\ref{1}) the exchange of states is not a conditional
operation, whereas the case stated above in \textsc{Defition 1} is
a conditional operation where the condition is on the global state
of the bipartite system. Aside from be distinct of a conventional
controlled operation where the condition is on a single part which
plays the roll of a control qubit. This result constitute a
counterpart of the conditional quantum dynamics stated in
reference \cite{barenco}, where one partie of a composite system
undergoes a coherent evolution that depends on the state of
another partie. In the present case, we will obtain \emph{a
conditional quantum dynamics in which a two parties system
undergoes a coherent evolution that depends on the overall state
of the whole system}. In the former case we can thought the
conditional evolution as $\hat{U}^{i}|i\rangle|j\rangle$, in the
latter case the conditional evolution is
$\hat{U}^{\left|i-j\right|}|i\rangle|j\rangle$, where $\hat{U}$ is
a unitary evolution operator, which satisfy
$\hat{U}^{-1}=\hat{U}^{\dag}$. In fact, there are another
conditional two-qubits gate whose conditional evolution depends on
the global state of the whole system, see the relative phase gate,
$\hat{U}_{Phase}^{relative}$, in reference \cite{paulo1}.

From the conditional \textsc{Definition 1} of SWAP gate enunciated,
it operates on two qubit computational basis as follows:
\begin{eqnarray}
\hat{U}_{SWAP}\left\vert 0\right\rangle _{1}\left\vert
0\right\rangle _{2} &=&\left\vert 0\right\rangle _{1}\left\vert
0\right\rangle _{2},  \nonumber
\\
\hat{U}_{SWAP}\left\vert 1\right\rangle _{1}\left\vert 1\right\rangle _{2}
&=&\left\vert 1\right\rangle _{1}\left\vert 1\right\rangle _{2},  \nonumber
\\
\hat{U}_{SWAP}\left\vert 0\right\rangle _{1}\left\vert
1\right\rangle _{2}
&=&\hat{U}_{NOT}|0\rangle_1\hat{U}_{NOT}|1\rangle_2=\left\vert
1\right\rangle _{1}\left\vert 0\right\rangle _{2},  \nonumber
\\
\hat{U}_{SWAP}\left\vert 1\right\rangle _{1}\left\vert
0\right\rangle _{2}
&=&\hat{U}_{NOT}|1\rangle_1\hat{U}_{NOT}|0\rangle_2=\left\vert
0\right\rangle _{1}\left\vert 1\right\rangle _{2}.  \label{2}
\end{eqnarray}%
This gate is reversible in the sense that the output is a unique
function of
the input \cite{divincenzo}. The main difference with the equation (\ref{1}%
), is that the physical implementation does not require a swapping on the
states but it requires a fliping process.

Secondly, if we begin with two unknown qubits and apply the equation (\ref{2}%
), we obtain:

\begin{eqnarray}
&&\hat{U}_{SWAP}\left( \alpha _{1}\left\vert 0\right\rangle _{1}+\beta
_{1}\left\vert 1\right\rangle _{1}\right) \left( \alpha _{2}\left\vert
0\right\rangle _{2}+\beta _{2}\left\vert 1\right\rangle _{2}\right)
\nonumber \\
&=&\left( \alpha _{2}\left\vert 0\right\rangle _{1}+\beta
_{2}\left\vert 1\right\rangle _{1}\right) \left( \alpha
_{1}\left\vert 0\right\rangle _{2}+\beta _{1}\left\vert
1\right\rangle _{2}\right).
\end{eqnarray}%
Therefore, by means of a flipping process we obtain a conditional
swapping between two unknown qubits.

On the other hand, \textsc{Definition 1} suggests the following
gate classification:

\begin{description}
    \item[Gate Class 1] In this class there is a conditional
    evolution of only one qubit of the whole system or its evolution can be implemented
     by the evolution of a single system, i. e. there is an application of an unitary operator $\hat{U}$
    only on a single qubit when the condition is satisfied, for example $|i\rangle_1\hat{U}|j\rangle_2$.
    Therefore, this class requires the manipulation of only a single
    qubit. The c-not, $U_{Phase}^{Relative}$  and $\hat{U}_{Phase}^{C}$ \cite{paulo1} gates belong to this class.

    \item[Gate Class 2] In this class there is a conditional
    evolution of both qubits of the whole system and its evolution can not be implemented by the evolution of a
    single system, i. e. there is an application of unitary operators to both
    subsystems when the condition is satisfied, for example $\hat{U}|i\rangle_1\hat{U}|j\rangle_2$. Therefore,
    this class requires the
    manipulation of two qubits. The SWAP (\textsc{Definition 1})
    and Double c-not gates \cite{collins} belong to this class.
\end{description}

In this classification, we consider that the whole system's
evolution can be implemented by a single system evolution when
$\hat{U}|i\rangle_1\hat{U}|j\rangle_2=|i\rangle_1\hat{U}|j\rangle_2$.
For example,
$\hat{U}_{Phase}^{one-qubit}|i\rangle_1\hat{U}_{Phase}^{one-qubit}|j\rangle_2=
|i\rangle_1\left(\hat{U}_{Phase}^{one-qubit}|j\rangle_2\right)
\quad or \quad
\left(\hat{U}_{Phase}^{one-qubit}|i\rangle_1\right)|j\rangle_2 $;
where $\hat{U}_{Phase}^{one-qubit}|s\rangle=e^{is\phi}|s\rangle
,\quad s=0,1$, gives a phase $\phi$. On the other hand, an example
of overall system's evolution that can not be simulated by a
single system evolution is
$\hat{U}_{NOT}|i\rangle_1\hat{U}_{NOT}|j\rangle_2\neq
|i\rangle_1\left(\hat{U}^{one-qubit}|j\rangle_2\right) \neq
\left(\hat{U}^{one-qubit}|i\rangle_1\right)|j\rangle_2$, i. e. you
can not find one-qubit gate that could satisfy a equality for this
equation, precisely because you need to manipulate both
subsystems.

At this stage, we want to remark the usefulness of this result on
quantum computation. As it is well known, the practical problems of
construction of a quantum computer can be solved by means of a
distributed quantum computer, i.e. a quantum communication network
in wich each node can act as a sender or receiver and contains only
a small number of qubits \cite{nielsen}. These results guided to the
necessity to establish optimal implementations of nonlocal gates
only using local operations and classical communication (LOCC) and
shared entanglement \cite{eisert,guo1}.

For instance A. Barenco et. al. \cite{barenco} proposed the first
implementation of a nonlocal SWAP gate on two unknown qubits using
two shared ebits and four bits of classical communications
(proposed as the fourth application of the c-not gate in reference
\cite{barenco}). A slightly different implementation of the
nonlocal SWAP gate was proposed by L. Vaidman
\cite{vaidman,vaidman2}. Independently, D. Collins et. al.
\cite{collins} and J. Eisert et. al. \cite{eisert} have
demonstrated that the nonlocal implementation of the SWAP gate
requires as necessary and sufficient resources four bits of
classical communication together with two shared ebits, besides to
shown that a nonlocal c-not gate consumes one bit of classical
comunications and one shared ebit. A similar work regarding the
amount of separability was made by A. Chefles et. al.
\cite{chefles}. Also, Zheng et. al. \cite{zheng}, shown a protocol
to implement the nonlocal SWAP operator on two entangled pairs. An
additional study of the resources needed to implement nonlocal
operators was carried out by K. Hammerer et al. \cite{hammerer}
and by J. I. Cirac et. al. \cite{cirac}. These protocols have
reached the experimental realization \cite{guo,huelga}. Now, the
gate's classification given above and Definition 1, allows us to
explain why the SWAP gate uses more amount of resources than
others. In the following we tailors this explanation.

Eisert et. al. \cite{eisert} have established a protocol to
realize an arbitrary control-U gate where the unitary operator
$\hat{U}$ is applied to the target qubit. If we analyze this
protocol from the point of view of gate's classification and
\textsc{Definition 1}, the control-U gate belongs to \textbf{Gates
Class 1} too. Following their protocol, we consider the case when
Alice holds the qubit $\left(\alpha|0\rangle_a+\beta|1\rangle_a
\right)$, Bob holds
$\left(\gamma|0\rangle_B+\delta|1\rangle_B\right)$ and they share
the ebit
$\left(|0\rangle_{A_1}|0\rangle_{B_1}+|1\rangle_{A_1}|1\rangle_{B_1}\right)$.
Then, the first steps of the protocol correlates the state of
system $A$ with the state of system $B_1$, in such a way that if
the state of $A$ is $|0\rangle_A$ (or $|1\rangle_A$) then the
state of $B_1$ is $|0\rangle_{B_1}$ (or $|1\rangle_{B_1}$), i. e.
the state of $A$ and $B_1$ is
$\left(|0\rangle_{A}|0\rangle_{B_1}+|1\rangle_{A}|1\rangle_{B_1}\right)$.
This correlation allows us to apply a unitary operator on state
$|i\rangle_B$ conditioned on the state $|i\rangle_{B_1}$, but as
$A$ and $B_1$ are correlated too, then the operator application is
realized as if a conditional quantum evolution between
$|i\rangle_A$ and $|i\rangle_B$ were existed. However, this
protocol does not enable to apply an unitary operator on the qubit
$|i\rangle_A$.

Therefore, sharing only one ebit and using two bits of classical
communications we can manipulate only one of the qubits of Alice and
Bob, because these resources are not sufficient to apply unitary
operators to both qubits. Therefore, the nonlocal implementation of
\textbf{Gates Class 2} uses more amounts of resources than the
\textbf{Gates Class 1} because in the former class it is necessary
to apply two unitary operators to both qubits of the whole system.

\hspace{2cm}

THREE SWAP GATE

\hspace{2cm}

 To analize the three qubits swap case, firstly we need to define what does mean the swapping
between three qubits. We propose that a three qubits swapping is
implemented as similar as a three axis rotation, that is:

\begin{eqnarray}
\hat{U}_{SWAP}^3\left(\alpha_1|0\rangle_1+\beta_1|1\rangle_1\right)
\otimes \left(\alpha_2|0\rangle_2+\beta_2|1\rangle_2\right)
\nonumber \\
\otimes \left(\alpha_3|0\rangle_3+\beta_3|1\rangle_3\right)
\nonumber\\=\left(\alpha_3|0\rangle_1+\beta_3|1\rangle_1\right)
\otimes \left(\alpha_1|0\rangle_2+\beta_1|1\rangle_2\right) \nonumber\\
\otimes \left(\alpha_2|0\rangle_3+\beta_3|1\rangle_3\right)
\end{eqnarray}

 However, a conditional definition similar to
\textsc{Definition 1} does not work, because it does not produce
the required change in the computational basis of three qubits.

\hspace{2cm}

 FREDKIN GATE

\hspace{2cm}

On the other hand, it is possible to construct a conditional
definition of the Fredkin gate \cite{nielsen,wang} as follows. The
Fredkin gate acts in the computational basis of three qubits,
where the first qubit plays the roll of control qubit.

\begin{quote}
\textsc{Definition 2}. If the control qubit is set to $|1\rangle$,
then apply a single qubit quantum NOT gate on both qubits if they
are not equal. Otherwise, left them unchanged.
\end{quote}

In conclusion, \textsc{Definition 1} can throw new light on the
seemingly paradoxical situation of uniquely using two ebits when
implementing a nonlocal SWAP gate compared with three ebits when
implementing the SWAP gate using three c-not gates.

\hspace{2cm}

\acknowledgments One of us, P. C. Garcia Quijas thanks the support
by Consejo Nacional de Ciencia y Tecnolog\'{\i}a (CONACYT). L. M.
Ar\'{e}valo Aguilar thanks to Sistema Nacional de Investigadores
(SNI) of Mexico.


\begin{thebibliography}{99}

\bibitem{barenco} Barenco, A., Deutsh, D., Ekert, A. \& Jozsa, R. Conditional quantum dynamics
and logic gates. \textit{Phys. Rev. Lett} \textbf{74}, 4083 (1995).

\bibitem{kumar} Kumar, A. \& Skinner, S. R. Simplified approach to implementing controlled-unitary
operations in a two-qubit system. \textit{Phys. Rev. A} \textbf{76},
022335 (2007).

\bibitem{paulo1} Garc\'{i}a Quijas, P. C. \& Ar\'{e}valo Aguilar, L. M. On the bandgap quantum coupler and the harmonic oscillator
interacting with a reservoir: Defining the relative phase gate.
arXiv:quant-ph/0702261v2.

\bibitem{feynman} Feynman, R. Quantum mechanical computers. \textit{Opt. News} \textbf{11} (2), 11 (1985).

\bibitem{deutsch} Deutsch, D., Quantum computational networks \textit{Proc. R. Soc. Lond. A} \textbf{425},
73 (1989).

\bibitem{nielsen} M. A. Nielsen and I. L. Chuang, Quantum Computation and
Quantum Information, Cambridge University Press (2000).

\bibitem{deutsch2} Deutsch, D., Barenco, A., \& Ekert, A. Universality in quantum computation.
\textit{Proc. R. Soc. Lond. A} \textbf{449}, 669 (1995).

\bibitem{nielsen-a} Nielsen, M. A. \& Chuang, I. L. Programmable quantum gate arrays.
\textit{Phys. Rev. Lett} \textbf{79}, 321 (1997).

\bibitem{vaidman} Vaidman, L. Teleportation of quantum states. \textit{Phys.
Rev. A} \textbf{49}, 1473 (1994).

\bibitem{vaidman2} Vaidman, L. \& Yoran, N. Methods for reliable teleportation. \textit{Phys.
Rev. A} \textbf{59}, 116 (1999).

\bibitem{brassard} Brassard, G., Braunstein, S. L., \& Cleve, R. Teleportation as quantum computation.
 \textit{Physica D} \textbf{120}, 43 (1998).

\bibitem{loss} Loss, D. \& DiVincenzo, D. P. Quantum computation
with quantum dots. \textit{Phys. Rev. A} \textbf{57}, 43 (1998).

\bibitem{eckert} Eckert, et. al. Quantum computing in optical microtraps based on the motional state of
neutral atoms. \textit{Phys. Rev. A} \textbf{66}, 43 (2002).

\bibitem{fan} Fan, H., Roychowdhury, V. \& Optimal two-qubit quantum circuits using exchange interactions.
\textit{Phys. Rev. A} \textbf{72}, 052323 (2005).

\bibitem{vatan} Vatan,F. \& Williams, C. Optimal quantum circuits for general two-qubits gates.
\textit{Phys. Rev. A} \textbf{69}, 032315 (2004).

\bibitem{abarenco} Barenco, A., et. al
Elementary gates for quantum computation, \textit{Phys. Rev. A}
\textbf{52}, 3457 (1995).

\bibitem{linden} Linden, N., Barjat, H., Kup\v{c}e, \={E}. \& Freeman, R. How to exchange information
between two coupled nuclear spins: the universal SWAP operation.
\textit{Chem. Phys. Lett.} \textbf{307}, 198 (1999).

\bibitem{fiorentino} Fiorentino, M., Kim, T., \& Wong, F. N. C. Single-photon two-qubit SWAP
gate for entanglement manipulation.  \textit{Phys. Rev. A}
\textbf{72}, 012318 (2005).


\bibitem{hardy} Hardy, Y., \& Steeb, W-H. Decomposing the SWAP quantum gate. \textit{J. Phys. A: Math. Gen.}
\textbf{39}, 1463 (2006).


\bibitem{divincenzo} DiVincenzo, D. P. Quantum gates and circuits. \textit{Proc. R. Soc. Lond. A} \textbf{454},
261 (1998).


\bibitem{collins} Collins, D., Linden, N. \& Popescu, S. Nonlocal content of quantum operations. \textit{Phys. Rev. A}
\textbf{64}, 032302 (2001).

\bibitem{eisert} Eisert, J., Jacobs, K., Papadopoulos, P. \& Plenio, M.
B. Optimal implementation of nonlocal quantum gates. \textit{Phys.
Rev. A} \textbf{62}, 052317 (2000).

\bibitem{guo1} Zhang, Yong-Sheng., Ye, Ming-Yong., \& Guo, Guang-Can.
Conditions for optimal construction of two-qubit nonlocal gates.
\textit{Phys. Rev. A} \textbf{71}, 062331 (2005).

\bibitem{chefles} Chefles, A., Gilson, R. C. \& Barnett, S. M.
Entanglement, information, and multiparticle quantum operatios.
\textit{Phys. Rev. A} \textbf{63}, 032314 (2001).

\bibitem{zheng} Zheng, Yi-Zhuang., Gu, Yong-Jian., \& Guo, Guang-Can.
Implementation of nonlocal quantum swap operation on two entangled
pairs \textit{Chin. Phys.} \textbf{11}, 529 (2002).

\bibitem{hammerer} Hammerer, K., Vidal, G. \& Cirac, J. I.
Characterization of nonlocal gates. \textit{Phys. Rev. A}
\textbf{66}, 062321 (2002).

\bibitem{cirac} Cirac, J. I., D\"{u}r, W., Kraus, B. \& Lewenstein, M.
Entangling operations and their implementation using small amount
of entanglement. \textit{Phys. Rev. Lett.} \textbf{86}, 544
(2001).

\bibitem{guo} Huang, Yun-Feng.,  Ren, Xi-Feng., Zhang, Yong-Sheng., Duan, Lu-Ming., \& Guo, Guang-Can.
Experimental Teleportation of a Quantum Controlled-NOT Gate.
\textit{Phys. Rev. Lett.} \textbf{93}, 240501 (2004).

\bibitem{huelga} Huelga, S. F., Plenio, M., Xiang, Guo-Yong., Li, Jian., \& Guo, Guang-Can.
Remote implementation of quantum operations. \textit{J. Opt. B:
Quantum Semiclass. Opt.} \textbf{7}, S384 (2005).

\bibitem{wang} Wang, B. \& Duan., L. M. Implementation scheme of controlled SWAP gates for quantum
fingerprinting and photonic quantum computation \textit{Phys. Rev.
A} \textbf{75}, 050304(R) (2007).

\end{thebibliography}
\end{document}